\documentclass[a4paper]{jpconf}
\usepackage{graphicx}
\usepackage[hidelinks]{hyperref}
\usepackage{listings}
\usepackage{subfig}
\usepackage{xcolor}

\definecolor{lbcolor}{rgb}{0.9,0.9,0.9}
\definecolor{cosmiclatte}{rgb}{1.0, 0.97, 0.91}     
\usepackage[sort&compress,numbers]{natbib}
\newcommand{\castro}{{\sf Castro}}
\newcommand{\amrex}{{\sf AMReX}}
\newcommand{\pynucastro}{{\sf pynucastro}}
\newcommand{\microphysics}{{\sf Microphysics}}

\usepackage[normalem]{ulem}

\usepackage{amsmath}
\usepackage{amssymb}

\begin{document}
\title{A Framework for Exploring Nuclear Physics Sensitivity in Numerical Simulations}

\author{Zhi Chen$^1$,
        Eric T. Johnson$^1$,
        Max Katz$^1$,
        Alexander Smith Clark$^1$,
        Brendan Boyd$^1$,
        Michael Zingale$^1$}

\address{$^1$Department of Physics and Astronomy, Stony Brook
  University, Stony Brook, NY 11794-3800 USA}

\ead{zhi.chen.3@stonybrook.edu}

\begin{abstract}
We describe the AMReX-Astrophysics framework for exploring
the sensitivity of astrophysical simulations to the details
of a nuclear reaction network, including the number of nuclei,
choice of reaction rates, and approximations used.  This is
explored by modeling a simple detonation with the \castro\
simulation code.  The entire simulation methodology is open-source and GPU-enabled.
\end{abstract}

\section{Introduction}

% Discuss the importance of nuclear reaction networks in simulating thermonuclear flames such as type 1a supernovae, x-ray bursts, nova, etc. 
% Discuss different networks available and used by others in the field.

% Give an overall background on how what part of astrophysics (perhaps even chemistry?) relies on nuclear reactions.

Nucleosynthesis is the process of forming new atomic nuclei via nuclear reactions. By studying both stellar and primordial nucleosynthesis, we gain better insights into stellar evolution, the chemical evolution of galaxies, and the element composition of the Universe. Moreover, enormous amounts of energy are released during nuclear fusion reactions in stellar nucleosynthesis, which is the power source leading to thermonuclear explosions such as Type Ia supernovae (SN Ia) \citep{Hillebrandt:2000, Maoz:2014}
%\citep{nomoto:1997, Lach:2020} 
and X-ray bursts (XRBs) \cite{galloway:2017}.
%\citep{parikh:2013} (more citations?)
Given the wide range of astrophysical environments in which nucleosynthesis occurs, each characterized by distinct thermodynamic conditions and dynamics, nuclear reaction networks are generally
tailored to each case.

% Explain the difficulty in including a limited amount of isotopes and reaction rates in the network
% How does things scale with dimension
%Numerical simulations are often complemented with observational data and theory to provide an accurate description of the system that we are studying. 
Due to limited computational resources, simulations of astrophysical
explosions often use small to moderate-sized networks.  Generally, the number of isotopes and reaction rates that can be included depends on the dimensionality of a simulation. In 1D, it is possible to incorporate more than a thousand isotopes along with thousands of nuclear reaction rates \citep{woosley-xrb}.
%(give citations, code, studies that do 1D reaction studies).
However, the situation becomes significantly more challenging in 2- and 3D, due to the increased memory and the need to advect each species separately.
%when the dimensionality of the simulation is increased (how does difficulty scale with dimension?) along with the coupling of other physics, such as hydrodynamics (perhaps give citation and the difficulty in hydro-reaction coupling).
A contemporary 2D XRB simulation can include 25 isotopes easily \cite{Chen_2023}, while a 3D XRB simulation can be restricted
to $\sim 10$ isotopes \citep{xrb3d, Zingale_2023}.

A reaction sensitivity study is essential to assess the reliability of a simulation where reactions provide a significant part of the energy budget.  This can explore the significance of each individual reaction rate, the size of the network, the modifications to rates (like screening), and approximations used in the network.  For example,
through a sensitivity study, we can identify the most critical reaction rates that should be included in the network while eliminating unimportant ones to reduce cost. During this process, we also gain insights into how reaction rates impact the properties of the problem, such as the propagating speed of a thermonuclear flame in an XRB \citep{Chen_2023} and the light curve of thermonuclear explosions, leading to a better convergence between the numerical models and observations.

% Discuss other reaction sensitivity studies in the field
%Various sensitivity studies have been done in the past to ...

% Present and give an outline of our framework

Here we present the \amrex-Astrophysics / \pynucastro\ framework for exploring nuclear reaction physics in numerical simulations.  The reaction networks are generated via the \pynucastro\ library \cite{pynucastro, pynucastro2}, which
writes GPU-enabled C++ networks for use in the \amrex-Astrophysics \microphysics\ library \cite{microphysics}. In section \ref{Section:generation}, we describe the overall procedure and workflow for generating an arbitrary reaction network with its right-hand-side (RHS) and Jacobian using \pynucastro. In section \ref{Section:testing}, we demonstrate a simple sensitivity study using the compressible hydrodynamics code \castro~\cite{castro,castro_joss} to model a 1D ${}^{4}\mathrm{He}$ detonation.  All of these codes are developed openly and are freely available on GitHub.

\section{Arbitrary Network Generation}
\label{Section:generation}
% Briefly include sample codes to include arbitrary rates from ReacLib library, tabulated weak rates, DerivedRates, Custom Rates in Pynucastro.
% Briefly discuss the ability to output rhs in both Python and C++.
% The overall workflow going from producing the rhs in Pynucastro. Nuclear burning in Microphysics using the rhs provided from the pynucastro script, and hydro coupling using Castro.
% Refer to the main pynucastro paper

% First Introduce Pynucastro
The first step in a reaction sensitivity study is the generation of the reaction network. We will use \pynucastro\ for this purpose. \pynucastro\ is a Python library that allows users to interactively explore nuclear reaction rates, construct any arbitrary reaction network, and export the RHS and Jacobian for the network for use in an integrator.  The {\tt ReacLibLibrary} class provides access to the 
collection of JINA ReacLib \citep{reaclib} rates and the 
{\tt TabularLibrary} class provides tabulated weak rates from
various sources \cite{Suzuki:2016,LANGANKE:2001}.  A detailed
description of \pynucastro\ can be found in \cite{pynucastro2}, here we only emphasize essential points related to network generation. 

% Show a sample network script and explain step by step.

To illustrate the process of network generation, we start by explaining the design philosophy and steps involved in building the {\tt subch\_simple} network, a network employed in our prior research studies \citep{Chen_2023, zingale2023sensitivity}. Subsequently, we showcase how this methodology can be seamlessly extended to develop more complex networks.

The first step in the network generation process is selecting the nuclei of interest.  We will build a network that is appropriate for
He burning in a detonation.  Both {\tt ReacLibLibrary} and {\tt TabularLibrary} provide a method called {\tt linking\_nuclei()} that takes a list of nuclei as input and returns a {\tt Library} object with all possible rates that
connect the input nuclei.   This is demonstrated in the following code:

\begin{lstlisting}
import pynucastro as pyna

# Create a library object that contains all ReacLib rates
reaclib_lib = pyna.ReacLibLibrary()

# Create a list of nuclei
nuc_list = ["p", "he4", "c12", "o16", "ne20", "mg24", "si28",
            "s32", "ar36", "ca40", "ti44", "cr48", "fe52", "ni56",
            "al27", "p31", "cl35", "k39", "sc43", "v47", "mn51",
            "co55", "n13", "n14", "f18", "ne21", "na22", "na23"]

# Returns a subset of ReacLib Library
subch_lib = reaclib_lib.linking_nuclei(nuc_list)
\end{lstlisting}

% Once you've established a base network by specifying all the involved nuclei, you may need to perform various refinement steps to reduce the network's overall size. It's often desirable to manually remove rates that have been determined as insignificant for energy generation or nucleosynthesis based on past studies. This can be accomplished effortlessly using the built-in {\tt remove\_rate()} method of the Library object. You can supply either the Rate object or the short name of the reaction rate in the form of ``A(x,y)B" as a string. Here's an example illustrating this functionality:

Next, various fine-tuning procedures can be done on this {\tt Library} to reduce the overall size of the network. It is often desirable to manually remove rates that have been determined as insignificant for energy generation or nucleosynthesis based on past studies. This can be accomplished using {\tt Library} built-in method {\tt remove\_rate()} by supplying either the {\tt Rate} object or the short name of the rate in the form of a string: {\tt "A(x, y)B"}.  Below we demonstrate
this by removing the reverse rates of C and O reactions as well as some links involving heavy nuclei:

\begin{lstlisting}
for r in subch_lib.get_rates():
    # Check if the products of the rate are the following
    if sorted(r.products) in [[pyna.Nucleus("c12"), pyna.Nucleus("c12")],
                              [pyna.Nucleus("c12"), pyna.Nucleus("o16")],
                              [pyna.Nucleus("o16"), pyna.Nucleus("o16")]]:
        # Remove the rate
        subch_lib.remove_rate(r)

# A list of rates to remove by simply supplying the short-name of the rate
rates_to_remove = ["p31(p,c12)ne20", "si28(a,c12)ne20",
                   "ne20(c12,p)p31", "ne20(c12,a)si28",
                   "na23(a,g)al27", "al27(g,a)na23",
                   "al27(a,g)p31", "p31(g,a)al27"]

# Remove rates individually from the network
for r in rates_to_remove:
    subch_lib.remove_rate(r)
\end{lstlisting}

% Should we show code or just quickly go through and refer back to pynucastro paper.

We can also directly modify the products of a reaction, to further simplify the network.  This is useful, for example, in a neutron-capture sequence like A(B, n)Y(n, $\gamma$)C where the neutron capture on Y is significantly faster than the first reaction.  In this case, a simple approximation involves modifying the product of the first rate to yield the final product of the two-reaction sequence, C, while retaining the rate of the first reaction and adjusting the overall Q-value accordingly. This is demonstrated in the following code:

\begin{lstlisting}
# Select the reaction rates and end product to modify
modify_rates = [("c12(c12,n)mg23", "mg24"),
                ("o16(o16,n)s31", "s32"),
                ("o16(c12,n)si27", "si28")]

for r, mp in modify_rates:
    # Get the specifc Rate object using the short-name
    _r = reaclib_lib.get_rate_by_name(r)
    # Modify the product of that rate
    _r.modify_products(mp)
    # Add the modified rate into the customized Library.
    subch_lib.add_rate(_r)
\end{lstlisting}

This network describes $\alpha$-captures up to the iron
group (and is the starting point for the {\tt subch\_simple} network from \cite{Chen_2023}).  However, for
modeling SN Ia, the formation of iron-peak isotopes via electron capture during the explosion is important \citep{nomoto:1997, parikh:2013, Bravo:2019, Lach:2020}.  We can include these by adding tabulated weak rates.   The code below obtains a new set of rates from both {\tt ReacLibLibrary} and  {\tt TabularLibrary} in the iron-peak and combines them with our existing rates to create our custom rate library {\tt all\_lib}. Note that while the original set of nuclei did not include neutrons, 
we do include neutrons in the iron-group reactions.

\begin{lstlisting}
# Define a new set of nuclei describing iron peak rates
iron_peak = ["n", "p", "he4", "mn51", "mn55",
             "fe52", "fe53", "fe54", "fe55", "fe56",
             "co55", "co56", "co57", "ni56", "ni57", "ni58",
             "cu59", "zn60"]

# Get a subset library with reaclib rates associated with these nuclei
iron_reaclib = reaclib_lib.linking_nuclei(iron_peak)

# Create the library that contains tabulated weak rates from Langanke
# and create a subset library with the rates associated with these nuclei
weak_lib = pyna.TabularLibrary()
iron_weak_lib = weak_lib.linking_nuclei(iron_peak)

# Combine all three libraries together.
all_lib = subch_lib + iron_reaclib + iron_weak_lib
\end{lstlisting}

There may be duplicated weak rates from the ReacLib library and the tabulated weak rates. To avoid errors when creating the network, these duplicates can be identified and removed via the {\tt find\_duplicated\_links()} method:

\begin{lstlisting}
# Find the duplicated links
dupes = all_lib.find_duplicate_links()

rates_to_remove = []
# Remove the ReacLibRate version in duplicates
for d in dupes:
    for r in d:
        if isinstance(r, pyna.rates.ReacLibRate):
            rates_to_remove.append(r)

# Remove rates
for r in rates_to_remove:
    all_lib.remove_rate(r)
\end{lstlisting}

% write out network

After we've completed the construction of the customized rate libraries, \pynucastro\ can output functions to compute both the righthand-side (RHS) of the ODE system of evolution equations and the Jacobian matrix.
Several formats are possible, each provided by a different class: {\tt PythonNetwork} for Python, {\tt SimpleCxxNetwork} for a basic C++ network, and {\tt AmrexAstroCxxNetwork} for C++ for use in the \amrex-Astrophysics suite. 
%In \cite{pynucastro2}, we've mainly demonstrated the applicability of solving the RHS generated from {\tt PythonNetwork} under a constant temperature and density using {\tt solve\_ivp()} from the scipy package. 
For our simulations, we want an {\tt AmrexAstroCxxNetwork}, which
can be used with \castro, and couples with hydrodynamics through a variety of integration techniques (e.g., our simplified-SDC method \cite{castro_simple_sdc}). We create a network in this format
simply as:

\begin{lstlisting}
# Use previously defined customized library to create our network
net = pyna.AmrexAstroCxxNetwork(libraries=[all_lib])
\end{lstlisting}

Finally, \pynucastro\ can also approximate rate sequences like A$(\alpha,\mbox{p})$X(p,$\gamma$)B rates into an effective as A$(\alpha, \gamma)$B rate, which is used in the traditional ``aprox'' \citep{timmes_aprox13} networks. This approximation holds valid when the proton-capture rate happens at a much faster timescale compared to the ($\alpha$, p) rate, causing an overall equilibrium in the proton flow. Similar to {\tt modify\_products()}, this approximation reduces the size of the network by removing the intermediate nuclei. This method is demonstrated in the following code:

\begin{lstlisting}
# Perform the (a,p)(p,g) approximation by specifying the intermediate nuclei
net.make_ap_pg_approx(intermediate_nuclei=["cl35", "k39", "sc43", "v47"])])

# Remove intermediate nuclei since now they're approximated out
net.remove_nuclei(["cl35", "k39", "sc43", "v47"])])
\end{lstlisting}

\begin{figure}[t]
    \centering
    \includegraphics[width=\linewidth]{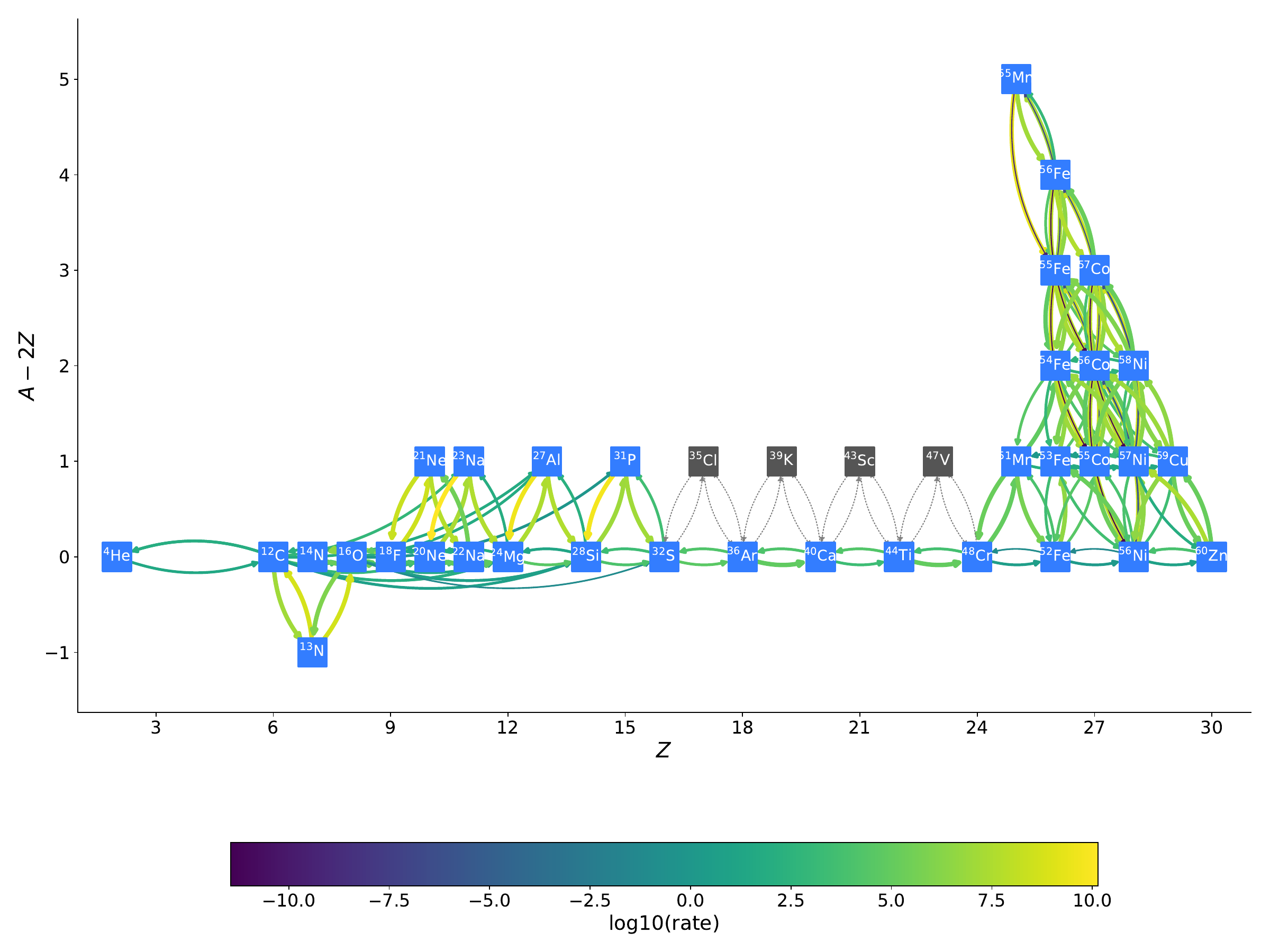}
    \caption{A visualization of the custom network that we created using the \pynucastro\ package. The color bar shows the reaction rates with solar composition, $\rho = 9 \times 10^7$ g $\mathrm{cm}^{-3}$, and $T = 6 \times 10^9$ K. The horizontal axis and the vertical axis show 
    the atomic number, $Z$, and the extra number of neutrons compared to the number of protons, respectively. The gray nodes with dotted gray lines represent the $(\alpha, \mbox{p})(\mbox{p}, \gamma)$ approximation, which are not directly in the network. All reactions that have the form, $A(X, \alpha)B$ or $A(\alpha, X)B$, are hidden for better clarity. }
    \label{fig:network_plot}
\end{figure}

Our final network has 36 isotopes and 149 rates including 12 {\tt TabularRate} rates (compared to the original {\tt subch\_simple} with 22 isotopes and 57 rates). Note that {\tt subch\_simple} also uses the ($\alpha$, p)(p, $\gamma$) approximation for ${}^{35}$Cl, ${}^{39}$K, ${}^{43}$Sc, ${}^{47}$V, ${}^{51}$Mn, and ${}^{55}$Co. Before writing out the actual RHS and the Jacobian, there are also methods that allow the user to explore the network they created and make adjustments if needed. For example, there is an easy interface to plot each reaction rate with a specified density, temperature, and mass abundances that mimic the simulation conditions to help determine the importance of each rate. An example output of the built-in plotting functions is shown in Figure \ref{fig:network_plot}.

% this will be in Alex's paper
%The {\tt find\_unimportant\_rates()} method within all Network classes provides an automatic way of identifying reaction rates under a set of conditions that are smaller than some user-defined percentage of the fastest rate in the current network. A network reduction algorithm based on pyMARS \citep{pyMARS} is also implemented recently to automatically eliminate unnecessary nuclei in the network to help trim down the network size. Examples of these can be found in the online documentation for \pynucastro.

Once we're satisfied with the current network, \pynucastro\ can write out all the necessary files, including the RHS and the Jacobian of the network, compatible with \microphysics\ and \castro\ with a single function:

\begin{lstlisting}
# Write all files needed by the package, Microphysics, to integrate the network
net.write_network()
\end{lstlisting}

The \amrex-Astrophysics \microphysics\ library contains
all of the code needed to support the integration of this network, including an equation of state (we'll use the Helmholtz EOS from \citep{timmes_swesty:2000}), integrators such as VODE \citep{vode}, as well as routines for neutrino cooling and plasma screening.  \microphysics\ can be used on its own to perform network evolution with self-heating or coupled to a simulation code, like \castro.  It provides
the necessary interfaces to support several different methods of coupling hydrodynamics and reactions.

Typically, a user creates a dedicated directory in \microphysics\
under {\tt Microphysics/networks} (we'll use {\tt custom\_net} for the network described above).  This directory serves as the repository containing both the network-generating script using \pynucastro\ and all the associated auto-generated files. 

% Introduce capability to evolve any arbitrary network in both c++ and python.

% Give samples codes of generating any arbitrary network, perhaps 2 or 3 different networks from the simplest

% Showcasing that we can directly use scipy to evolve the system but doesn't have the eos yet so segway to microphysics so that we can also evolving temperature and density.

\section{Network Sensitivity Testing}
\label{Section:testing}
% Maybe combine this section with previous section
% This section primarily showcasing microphysics, maybe show burncell,
% Introduce microphysics, and how rhs from pynucastro can be used very easily.

% Give some words about pure c++ network, where there is no amrex dependence

We will now show how to do a network sensitivity study similar to
the one recently performed for XRBs in \cite{Chen_2023}.  Any
\castro\ problem setup can be told to use our new network
(in the \microphysics\ directory hierarchy) at compile time, via:

\begin{lstlisting}
make NETWORK_DIR=custom_net
\end{lstlisting}

%A simple demonstration of this self-heating evolution is the {\tt burn\_cell} unit test in \microphysics. Given the initial conditions in density, temperature, and initial mass abundance of the species, {\tt burn\_cell} evolves the reaction network over some user-defined time in a single cell and records the evolving states. However, a basic self-heating reaction network is inadequate when it comes to investigating scientific phenomena like thermonuclear explosions, where the interplay between hydrodynamics and nuclear burning is important. Therefore, we use \castro, a compressible hydrodynamics simulation code with adaptive mesh refinement, in conjunction with \microphysics\ to achieve a reaction-hydrodynamics evolution. 

To demonstrate a simple sensitivity testing, we use the {\tt Detonation} test problem in \castro. {\tt Detonation} models a 1D propagating detonation wave, offering a way to replicate the thermodynamic conditions of the He detonation wave found in the shell-burning stage of the sub-Chandrasekhar double-detonation model \cite{fink:2007}, as discussed in previous work \citep{castro_simple_sdc, zingale2023sensitivity}. This sensitivity testing on the {\tt Detonation} problem helps us choose the appropriate network for a realistic sub-Chandrasekhar double-detonation model, especially when studying nucleosynthesis.

To demonstrate this idea, we will compare {\tt subch\_simple} and the classic 19-isotope network, {\tt aprox19} \citep{Kepler}, with the network we generated in Section \ref{Section:generation}, which will be referred to as {\tt He-C-Fe}.  Our detonation
simulation spans a domain up to $x = 4.0 \times 10^7$ cm, with 256 coarse grid zones. By employing two refinement levels with ratios of 2, we reach 1024 zones at the finest level, providing a resolution of  $\sim 0.4$ km. A zero-gradient outflow boundary condition is used on both boundaries. The initial thermodynamic conditions are taken
from the base of the He layer from the simulations in  \cite{zingale2023sensitivity}: a uniform $\rho = 1.1 \times 10^9$ $\mbox{g}~\mbox{cm}^{-3}$, $\mbox{T}_{\mbox{pert}} = 1.1 \times 10^9$ K from $0 < x_{\mbox{pert}} < 1.2 \times 10^7$ cm,  $\mbox{T} = 1.75 \times 10^8$ K for $x > 1.2 \times 10^7$ cm, and a composition
of 99 \% ${}^{4}$He and 1 \% ${}^{14}$N.

\begin{figure}
    \centering
    \includegraphics[width=0.9\linewidth]{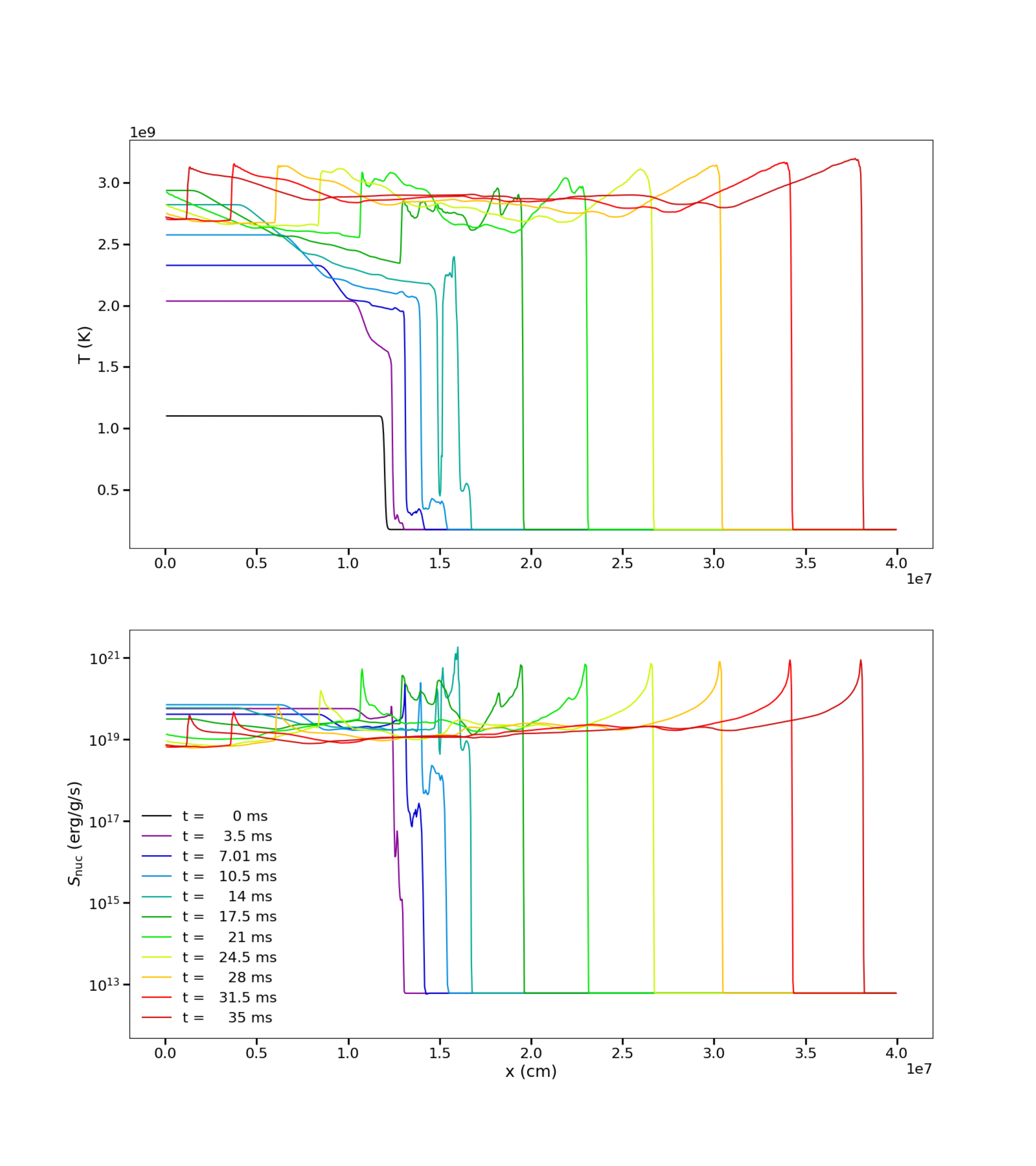}
    \caption{Temperature and specific nuclear energy generation evolution for {\tt He-C-Fe}}
    \label{fig:He-C-Fe_Te}
\end{figure}

Figure \ref{fig:He-C-Fe_Te} shows the time evolution of the temperature and specific nuclear energy generation rate using the {\tt He-C-Fe} network. The peak temperature and specific energy generation rate reach $\sim 3\times 10^9$ K and $\sim 10^{21}~\mbox{erg}~\mbox{g}^{-1}~\mbox{s}^{-1}$, consistent with results of the shell-burning stage in the previous study \citep{zingale2023sensitivity}.

\begin{figure}
    \centering
    \includegraphics[width=0.9\linewidth]{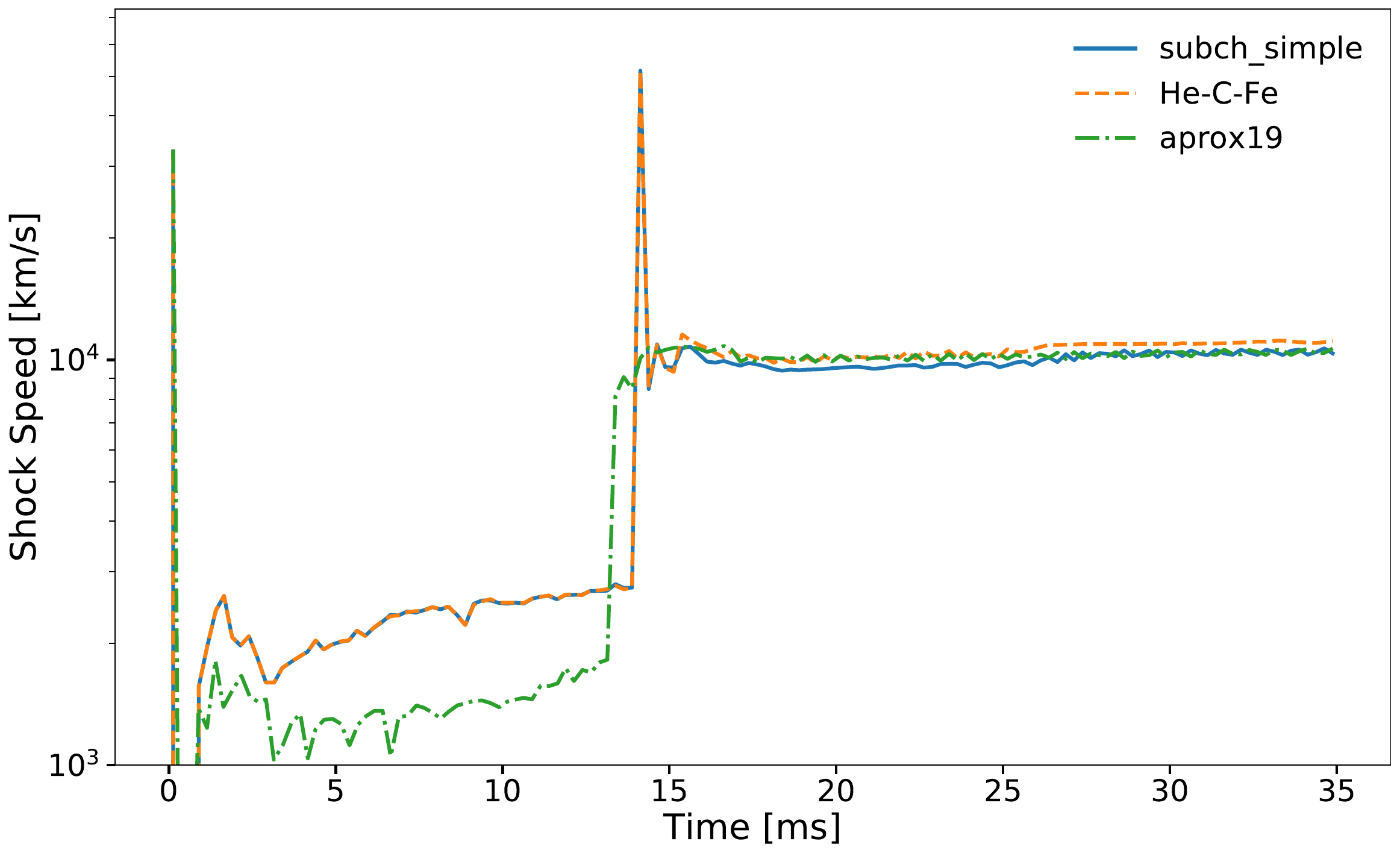}
    \caption{Time evolution of the shock speed for different networks.}
    \label{fig:speed}
\end{figure}

We now look at the effect of the network.  Figure \ref{fig:speed} shows the time evolution of the shock speed for our different networks. It is observed that all models reach a shock speed of $\sim 10^4$ $\mbox{km} \ \mbox{s}^{-1}$ after $\sim 15$ ms, while some noticeable differences for $t \lesssim 15$ ms. Specifically, {\tt He-C-Fe} and {\tt subch\_simple} exhibit similar profiles, with roughly double the shock speed compared to {\tt aprox19} for $t \lesssim 15$ ms. This difference in shock speed is due to the missing ${}^{12}\mbox{C}(\mbox{p}, \gamma) {}^{13}\mbox{N}(\alpha, \mbox{p}){}^{16}\mbox{O}$ rate in {\tt aprox19}, resulting in weaker initial carbon-burning \citep{Weinberg_2006, fisker:2008, Shen_2009}.  A similar effect was seen with flames
in XRBs in the study by \cite{Chen_2023}.

\begin{figure}
    \centering
    \subfloat[{\tt He-C-Fe}. Both solid and dashed lines are used to better distinguishability.]{\includegraphics[width=\linewidth]{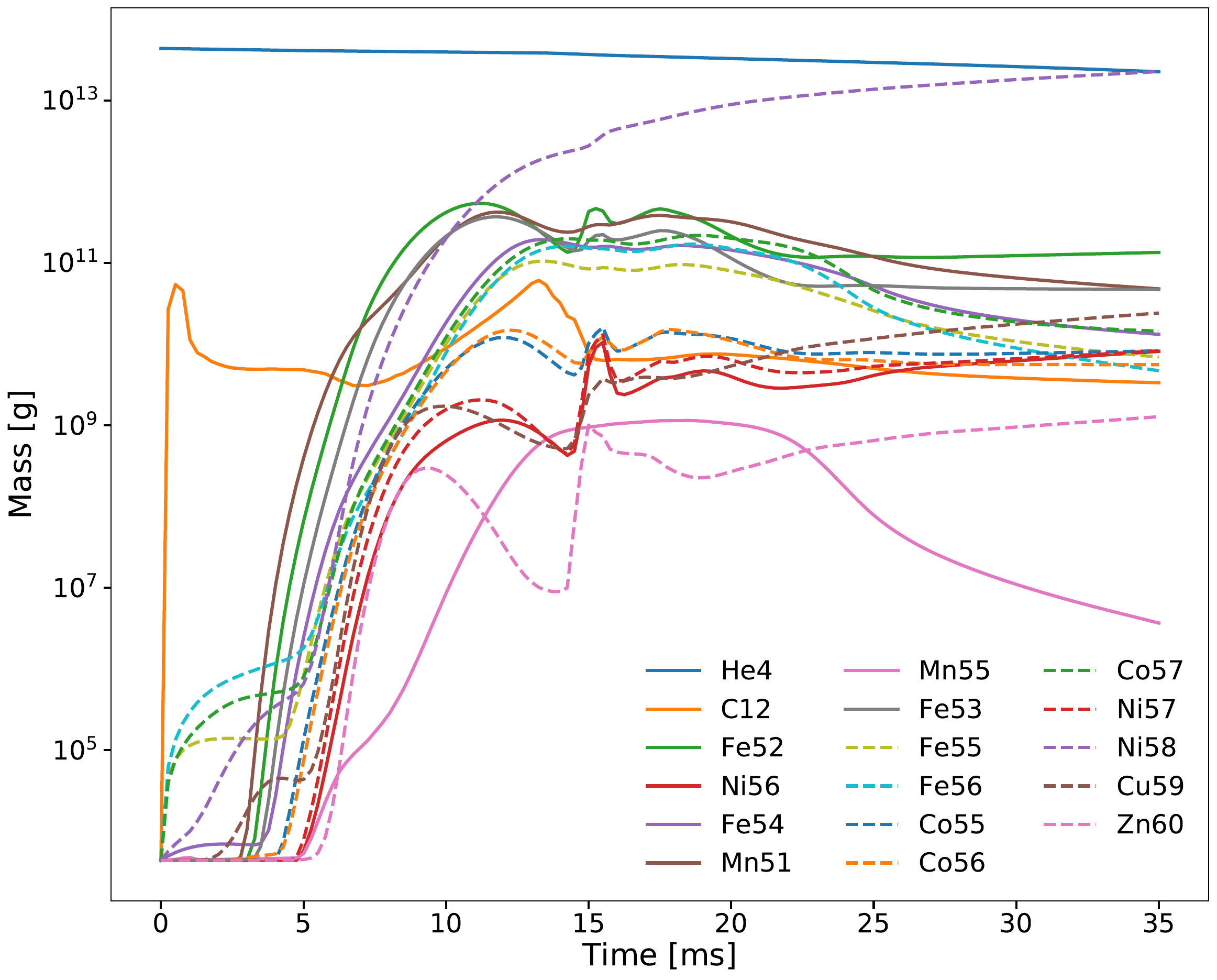}\label{fig:He-C-Fe_mass}}
    \hfill
    \subfloat[{\tt subch\_simple}]{\includegraphics[width=0.5\linewidth]{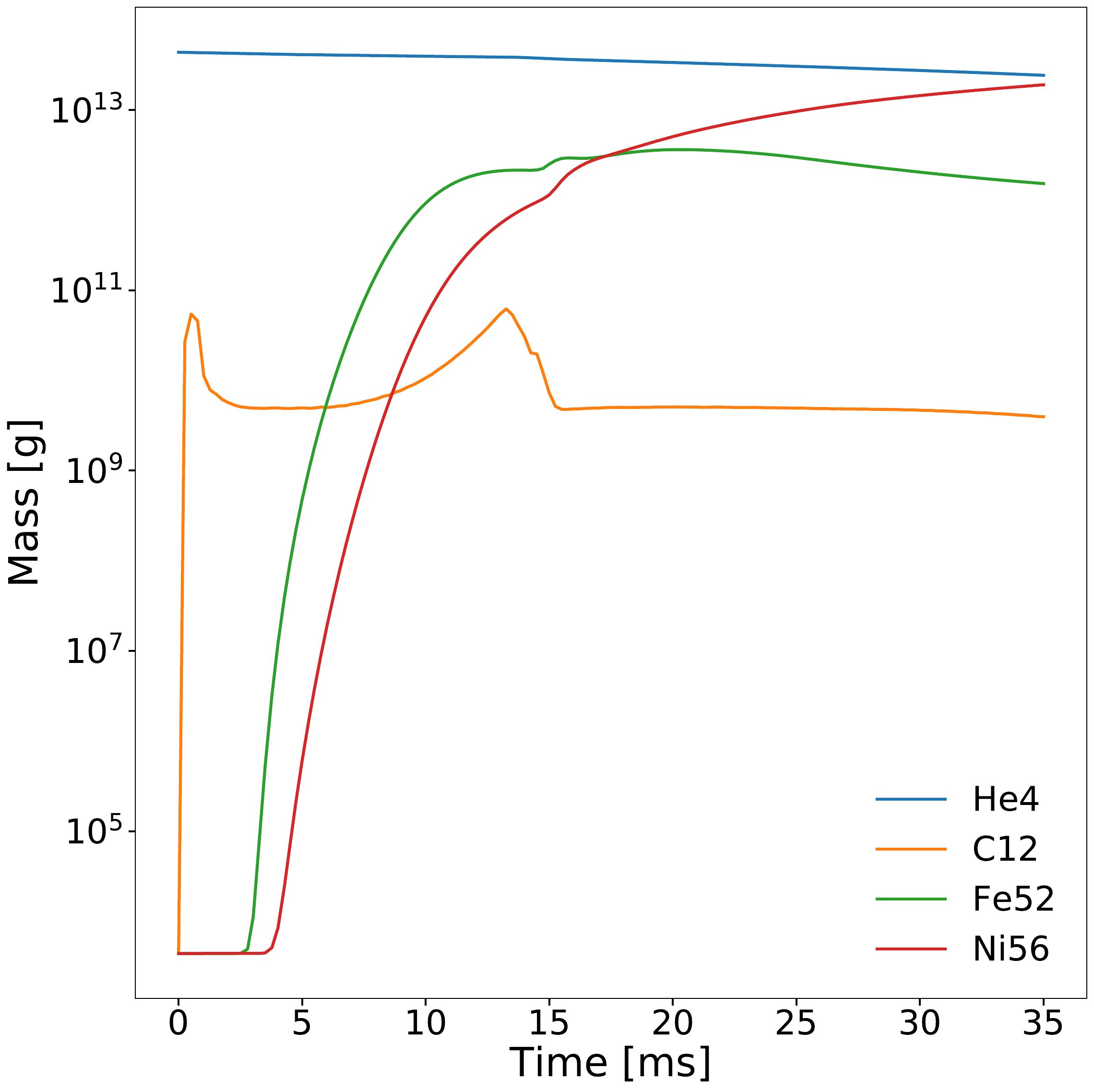}\label{fig:subch_simple_mass}}
    \hfill
    \subfloat[{\tt aprox19}] {\includegraphics[width=0.5\linewidth]{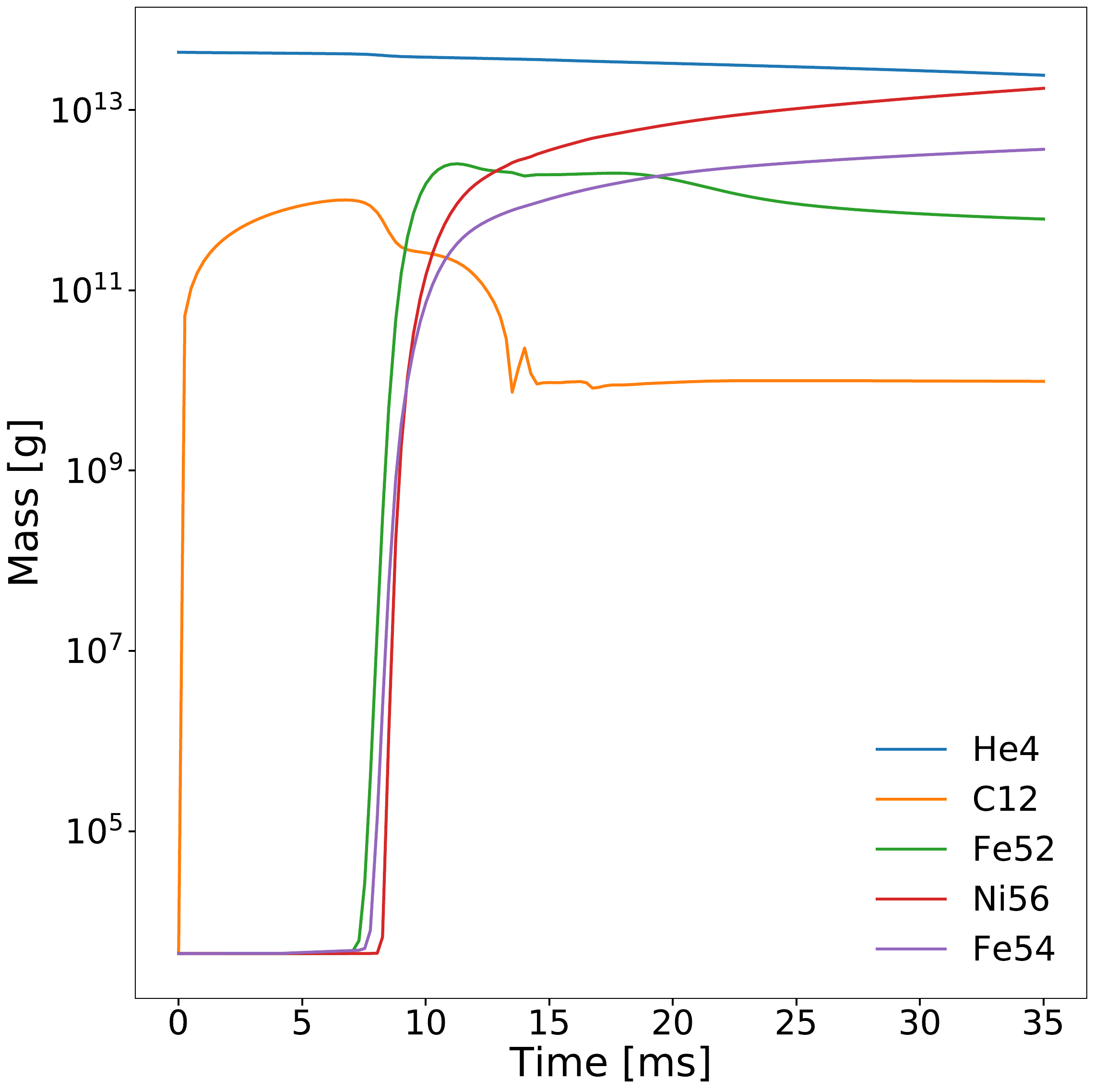}\label{fig:aprox19_mass}}
    \caption{\label{fig:mass}Time evolution of the total mass of ${}^{4}$He, ${}^{12}$C and iron-peak isotopes over time.}
\end{figure}

% Discuss networks used a bit
% Figure \ref{fig:mass} shows the time evolution of ${}^{4}$He, ${}^{12}$C, and iron-peak isotopes available in the network to examine the overall nucleosynthesis for the different networks. It is also observed that there is a direct correlation between the evolution of shock speed in Figure \ref{fig:speed} and the total mass abundance of ${}^{12}$C. Due to the enhanced carbon burning path that both {\tt He-C-Fe} and {\tt subch\_simple} have, more ${}^{12}$C are consumed leading to a higher shock speed. As the total mass of ${}^{12}$C stabilizes after $\sim 15$ ms, a steady speed is observed for all three networks.

This carbon burning enhancement effect is evident when examining the total mass abundance of ${}^{12}$C in Figure \ref{fig:mass}, showing the time evolution of ${}^{4}$He, ${}^{12}$C, and iron-peak isotopes available in the network. The enhanced carbon burning path in both {\tt He-C-Fe} and {\tt subch\_simple} leads to a greater ${}^{12}$C consumption rate, leading to a higher shock speed. After $\sim 15$ ms, the total mass of ${}^{12}$C stabilizes, reaching a steady speed for all three networks. In terms of iron-peak nucleosynthesis, both {\tt aprox19} and {\tt subch\_simple} indicate that ${}^{56}$Ni is the primary product, whereas {\tt He-C-Fe} predominantly yields ${}^{58}$Ni. This insight is important for determining the correct mass abundance of ${}^{56}$Ni following the explosion since the beta decay from ${}^{56}\mbox{Ni} \rightarrow {}^{56}\mbox{Co} \rightarrow {}^{56}\mbox{Fe}$ primarily shapes the observed SN Ia light curve.

% Discuss the results and show difference
% Nucleosynthesis results
% flamespeed 
% cpu-time used

\begin{figure}
    \centering
    \includegraphics[width=0.9\linewidth]{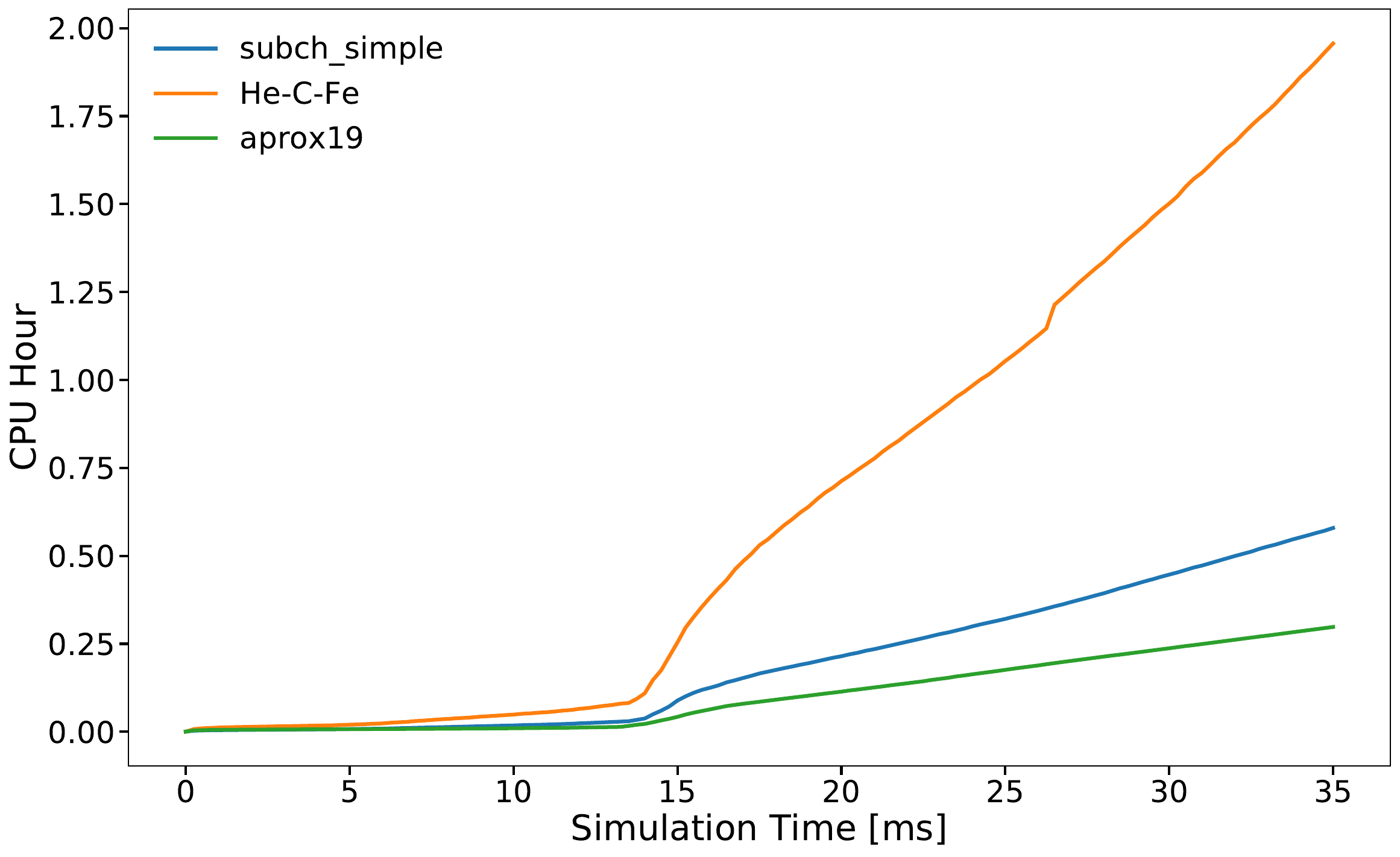}
    \caption{Total number of CPU hours used for each network.}
    \label{fig:cpu}
\end{figure}

To demonstrate the computational cost associated with adding additional iron-peak isotopes, Figure \ref{fig:cpu} shows the CPU-hours used with each network. As expected, a larger network size corresponds to a higher CPU hour usage. Specifically, the CPU-hours used by {\tt He-C-Fe} increased drastically after $15$ ms, coinciding with the establishment of the steady propagating detonation wave. In the end, {\tt He-C-Fe} required nearly 4 times the CPU-hours compared to {\tt subch\_simple}, which has  $\sim 1/3$ of the total number of rates in {\tt He-C-Fe}. This result suggests the need for further simplification and optimization of the network before employing it in actual simulations such as a 2D sub-Chandrasekhar double detonation model.

\section{Future Development}
% Discuss the ongoing effort to create a larger network for subchandra.

We've outlined the versatile approach for nuclear reaction sensitivity testing using \pynucastro, \microphysics, and \castro. By leveraging \pynucastro\ for network generation, \microphysics\ for network burning, and \castro\ for reaction-hydrodynamics coupling, we can effortlessly investigate any network of choice. Demonstrated through the 1D {\tt Detonation} problem in \castro, we assess reaction network sensitivity and emulate shell-burning of the sub-Chandrasekhar double detonation model nucleosynthesis. In the future, this testing capability aims to identify an efficient network with sufficiently many iron-peak isotopes, similar to {\tt He-C-Fe}, for studying nucleosynthesis in the actual sub-Chandrasekhar double detonation model.

\ack

\castro\ is open-source and freely available at
\url{http://github.com/AMReX-Astro/Castro}. The work at Stony Brook was supported by DOE/Office
of Nuclear Physics grant DE-FG02-87ER40317. This research was supported by the Exascale Computing 
Project (17-SC-20-SC), a collaborative effort of the U.S. Department of Energy
Office of Science and the National Nuclear Security Administration. This research used
resources of the National Energy Research Scientific Computing Center,
a DOE Office of Science User Facility supported by the Office of
Science of the U.~S.\ Department of Energy under Contract
No.\ DE-AC02-05CH11231.  This research used resources of the Oak Ridge
Leadership Computing Facility at the Oak Ridge National Laboratory,
which is supported by the Office of Science of the U.S. Department of
Energy under Contract No. DE-AC05-00OR22725, awarded through the DOE
INCITE program.  We thank NVIDIA Corporation for the donation of a
Titan X and Titan V GPU through their academic grant program.  This
research has made use of NASA's Astrophysics Data System Bibliographic
Services.

\bibliographystyle{iopart-num}
\bibliography{ws}

\providecommand{\newblock}{}
\begin{thebibliography}{10}
\expandafter\ifx\csname url\endcsname\relax
  \def\url#1{{\tt #1}}\fi
\expandafter\ifx\csname urlprefix\endcsname\relax\def\urlprefix{URL }\fi
\providecommand{\eprint}[2][]{\url{#2}}

\bibitem{Hillebrandt:2000}
Hillebrandt W and Niemeyer J~C 2000 {\em Annual Review of Astronomy and
  Astrophysics\/} {\bf 38} 191--230 (\textit{Preprint}
  \eprint{https://doi.org/10.1146/annurev.astro.38.1.191})
  \urlprefix\url{https://doi.org/10.1146/annurev.astro.38.1.191}

\bibitem{Maoz:2014}
Maoz D, Mannucci F and Nelemans G 2014 {\em Annual Review of Astronomy and
  Astrophysics\/} {\bf 52} 107--170 (\textit{Preprint}
  \eprint{https://doi.org/10.1146/annurev-astro-082812-141031})
  \urlprefix\url{https://doi.org/10.1146/annurev-astro-082812-141031}

\bibitem{galloway:2017}
{Galloway} D~K and {Keek} L 2021 {\em Timing Neutron Stars: Pulsations,
  Oscillations and Explosions\/} ({\em Astrophysics and Space Science
  Library\/} vol 461) ed {Belloni} T~M, {M{\'e}ndez} M and {Zhang} C pp
  209--262 (\textit{Preprint} \eprint{1712.06227})

\bibitem{woosley-xrb}
Woosley S~E, Heger A, Cumming A, Hoffman R~D, Pruet J, Rauscher T, Fisker J~L,
  Schatz H, Brown B~A and Wiescher M 2004 {\em ASTROPHYS J SUPPL S\/} {\bf 151}
  75--102 ISSN 0067-0049, 1538-4365
  \urlprefix\url{https://doi.org/10.1086/381533}

\bibitem{Chen_2023}
Chen Z, Zingale M and Eiden K 2023 {\em The Astrophysical Journal\/} {\bf 955}
  128 \urlprefix\url{https://dx.doi.org/10.3847/1538-4357/acec72}

\bibitem{xrb3d}
Zingale M, Malone C~M, Nonaka A, Almgren A~S and Bell J~B 2015 {\em ApJ\/} {\bf
  807} 60 ISSN 1538-4357 (\textit{Preprint} \eprint{1410.5796})
  \urlprefix\url{https://doi.org/10.1088/0004-637x/807/1/60}

\bibitem{Zingale_2023}
Zingale M, Eiden K and Katz M 2023 {\em ApJ\/} {\bf 952} 160
  \urlprefix\url{https://dx.doi.org/10.3847/1538-4357/ace04e}

\bibitem{pynucastro}
E~Willcox D and Zingale M 2018 {\em JOSS\/} {\bf 3} 588 ISSN 2475-9066
  \urlprefix\url{https://doi.org/10.21105/joss.00588}

\bibitem{pynucastro2}
Clark A~I~S, Johnson E~T, Chen Z, Eiden K, Willcox D~E, Boyd B, Cao L,
  DeGrendele C~J and Zingale M 2023 {\em ApJ\/} {\bf 947} 65
  \urlprefix\url{https://dx.doi.org/10.3847/1538-4357/acbaff}

\bibitem{microphysics}
{AMReX-Astro Microphysics Development Team}, Bishop A, Fields C~E, Chen Z,
  Harpole A, Jacobs A~M, Johnson E, Katz M, Li X, Malone C, Timmes F, Willcox D
  and Zingale M 2023 Amrex-astro/microphysics: Release 23.08
  \urlprefix\url{https://doi.org/10.5281/zenodo.8206742}

\bibitem{castro}
Almgren A~S, Beckner V~E, Bell J~B, Day M~S, Howell L~H, Joggerst C~C, Lijewski
  M~J, Nonaka A, Singer M and Zingale M 2010 {\em ApJ\/} {\bf 715} 1221--1238
  ISSN 0004-637X, 1538-4357 (\textit{Preprint} \eprint{1005.0114})
  \urlprefix\url{https://doi.org/10.1088/0004-637x/715/2/1221}

\bibitem{castro_joss}
Almgren A, Sazo M~B, Bell J, Harpole A, Katz M, Sexton J, Willcox D, Zhang W
  and Zingale M 2020 {\em JOSS\/} {\bf 5} 2513
  \urlprefix\url{https://doi.org/10.21105/joss.02513}

\bibitem{reaclib}
Cyburt R~H, Amthor A~M, Ferguson R, Meisel Z, Smith K, Warren S, Heger A,
  Hoffman R~D, Rauscher T, Sakharuk A, Schatz H, Thielemann F~K and Wiescher M
  2010 {\em ApJS\/} {\bf 189} 240--252 ISSN 0067-0049, 1538-4365
  \urlprefix\url{https://doi.org/10.1088/0067-0049/189/1/240}

\bibitem{Suzuki:2016}
Suzuki T, Toki H and Nomoto K 2016 {\em ApJ\/} {\bf 817} 163
  \urlprefix\url{https://dx.doi.org/10.3847/0004-637X/817/2/163}

\bibitem{LANGANKE:2001}
LANGANKE K and MARTÍNEZ-PINEDO G 2001 {\em Atomic Data and Nuclear Data
  Tables\/} {\bf 79} 1--46 ISSN 0092-640X
  \urlprefix\url{https://www.sciencedirect.com/science/article/pii/S0092640X01908654}

\bibitem{zingale2023sensitivity}
Zingale M, Chen Z, Rasmussen M, Polin A, Katz M, Clark A~S and Johnson E~T 2023
  Sensitivity of simulations of double detonation type ia supernova to
  integration methodology (\textit{Preprint} \eprint{2309.01802})

\bibitem{nomoto:1997}
{Nomoto} K, {Iwamoto} K, {Nakasato} N, {Thielemann} F~K, {Brachwitz} F,
  {Tsujimoto} T, {Kubo} Y and {Kishimoto} N 1997 {\em \nphysa\/} {\bf 621}
  467--476 (\textit{Preprint} \eprint{astro-ph/9706025})

\bibitem{parikh:2013}
Parikh A, José J, Sala G and Iliadis C 2013 {\em Prog. Part. Nucl. Phys.\/}
  {\bf 69} 225--253 ISSN 0146-6410 (\textit{Preprint} \eprint{1211.5900})
  \urlprefix\url{https://doi.org/10.1016/j.ppnp.2012.11.002}

\bibitem{Bravo:2019}
{Bravo, E} 2019 {\em A\&A\/} {\bf 624} A139
  \urlprefix\url{https://doi.org/10.1051/0004-6361/201935095}

\bibitem{Lach:2020}
{Lach, F}, {R\"opke, F K}, {Seitenzahl, I R}, {Cot\'e, B}, {Gronow, S} and
  {Ruiter, A J} 2020 {\em A\&A\/} {\bf 644} A118
  \urlprefix\url{https://doi.org/10.1051/0004-6361/202038721}

\bibitem{castro_simple_sdc}
Zingale M, Katz M~P, Nonaka A and Rasmussen M 2022 {\em ApJ\/} {\bf 936} 6
  \urlprefix\url{https://dx.doi.org/10.3847/1538-4357/ac8478}

\bibitem{timmes_aprox13}
{Timmes} F~X 2019 aprox13 network
  \urlprefix\url{https://cococubed.com/code\_pages/burn\_helium.shtml}

\bibitem{timmes_swesty:2000}
Timmes F~X and Swesty F~D 2000 {\em ApJS\/} {\bf 126} 501--516 ISSN 0067-0049,
  1538-4365 \urlprefix\url{https://doi.org/10.1086/313304}

\bibitem{vode}
Brown P~N, Byrne G~D and Hindmarsh A~C 1989 {\em SIAM J. Sci. and Stat.
  Comput.\/} {\bf 10} 1038--1051 ISSN 0196-5204, 2168-3417
  \urlprefix\url{https://doi.org/10.1137/0910062}

\bibitem{fink:2007}
{Fink} M, {Hillebrandt} W and {R{\"o}pke} F~K 2007 {\em \aap\/} {\bf 476}
  1133--1143 (\textit{Preprint} \eprint{0710.5486})

\bibitem{Kepler}
Weaver T~A, Zimmerman G~B and Woosley S~E 1978 {\em ApJ\/} {\bf 225} 1021 ISSN
  0004-637X, 1538-4357 \urlprefix\url{https://doi.org/10.1086/156569}

\bibitem{Weinberg_2006}
{Weinberg} N~N, {Bildsten} L and {Schatz} H 2006 {\em \apj\/} {\bf 639}
  1018--1032 (\textit{Preprint} \eprint{astro-ph/0511247})

\bibitem{fisker:2008}
Fisker J~L, Schatz H and Thielemann F 2008 {\em ApJS\/} {\bf 174} 261--276 ISSN
  0067-0049, 1538-4365 \urlprefix\url{https://doi.org/10.1086/521104}

\bibitem{Shen_2009}
Shen K~J and Bildsten L 2009 {\em ApJ\/} {\bf 699} 1365--1373
  \urlprefix\url{https://doi.org/10.1088%2F0004-637x%2F699%2F2%2F1365}

\end{thebibliography}

\end{document}